\documentclass{INTERSPEECH2023}

\usepackage{threeparttable}
\usepackage{subfiles}
\usepackage{multirow}
\usepackage{stfloats}
\usepackage{url}
\usepackage{colortbl}
\usepackage{xcolor}

\DeclareMathOperator{\Enc}{Enc}
\DeclareMathOperator{\CTC}{CTC}
\DeclareMathOperator{\Dec}{Dec}
\DeclareMathOperator{\Norm}{NRM}
\DeclareMathOperator{\Linear}{LIN}

\interspeechcameraready 

\title{A New Benchmark of Aphasia Speech Recognition and Detection Based on E-Branchformer and Multi-task Learning}
\name{Jiyang Tang$^1$, William Chen$^1$, Xuankai Chang$^1$, Shinji Watanabe$^1$, Brian MacWhinney$^2$}
\address{
  $^1$Language Technologies Institute, Carnegie Mellon University, USA\\
  $^2$Department of Psychology, Carnegie Mellon University, USA}
\email{jiyangta@cs.cmu.edu}

\usepackage[
backend=biber,
style=ieee,
citestyle=numeric-comp,
maxbibnames=3,
minbibnames=1,
maxcitenames=3,
doi=false,isbn=false,url=false,eprint=false
]{biblatex}

\defbibheading{bibliography}[\refname]{}

\DeclareSourcemap{
	\maps[datatype=bibtex, overwrite=true]{
		\map{
			\step[fieldsource=booktitle,
			match=\regexp{.*Interspeech.*},
			replace={Proc. Interspeech}]
			\step[fieldsource=journal,
			match=\regexp{.*INTERSPEECH.*},
			replace={Proc. Interspeech}]
			\step[fieldsource=booktitle,
			match=\regexp{.*ICASSP.*},
			replace={Proc. ICASSP}]
			\step[fieldsource=booktitle,
			match=\regexp{.*icassp_inpress.*},
			replace={Proc. ICASSP (in press)}]
			\step[fieldsource=booktitle,
			match=\regexp{.*Acoustics,.*Speech.*and.*Signal.*Processing.*},
			replace={Proc. ICASSP}]
			\step[fieldsource=booktitle,
			match=\regexp{.*International.*Conference.*on.*Learning.*Representations.*},
			replace={Proc. ICLR}]
			\step[fieldsource=booktitle,
			match=\regexp{.*International.*Conference.*on.*Computational.*Linguistics.*},
			replace={Proc. COLING}]
			\step[fieldsource=booktitle,
			match=\regexp{.*SIGdial.*Meeting.*on.*Discourse.*and.*Dialogue.*},
			replace={Proc. SIGDIAL}]
			\step[fieldsource=booktitle,
			match=\regexp{.*International.*Conference.*on.*Machine.*Learning.*},
			replace={Proc. ICML}]
			\step[fieldsource=booktitle,
			match=\regexp{.*North.*American.*Chapter.*of.*the.*Association.*for.*Computational.*Linguistics:.*Human.*Language.*Technologies.*},
			replace={Proc. NAACL}]
			\step[fieldsource=booktitle,
			match=\regexp{.*Empirical.*Methods.*in.*Natural.*Language.*Processing.*},
			replace={Proc. EMNLP}]
			\step[fieldsource=booktitle,
			match=\regexp{.*Association.*for.*Computational.*Linguistics.*},
			replace={Proc. ACL}]
			\step[fieldsource=booktitle,
			match=\regexp{.*Automatic.*Speech.*Recognition.*and.*Understanding.*},
			replace={Proc. ASRU}]
			\step[fieldsource=booktitle,
			match=\regexp{.*Spoken.*Language.*Technology.*},
			replace={Proc. SLT}]
			\step[fieldsource=booktitle,
			match=\regexp{.*Speech.*Synthesis.*Workshop.*},
			replace={Proc. SSW}]
			\step[fieldsource=booktitle,
			match=\regexp{.*workshop.*on.*speech.*synthesis.*},
			replace={Proc. SSW}]
			\step[fieldsource=booktitle,
			match=\regexp{.*Advances.*in.*neural.*information.*processing.*},
			replace={Proc. NeurIPS}]
			\step[fieldsource=booktitle,
			match=\regexp{.*Advances.*in.*Neural.*Information.*Processing.*},
			replace={Proc. NeurIPS}]
			\step[fieldsource=booktitle,
			match=\regexp{.*Workshop.*on.* Applications.* of.* Signal.*Processing.*to.*Audio.*and.*Acoustics.*},
			replace={Proc. WASPAA}]
			\step[fieldsource=publisher,
			match=\regexp{.+},
			replace={{}}]
			\step[fieldsource=month,
			match=\regexp{.+},
			replace={{}}]
			\step[fieldsource=location,
			match=\regexp{.+},
			replace={{}}]
			\step[fieldsource=address,
			match=\regexp{.+},
			replace={{}}]
			\step[fieldsource=organization,
			match=\regexp{.+},
			replace={{}}]
		}
	}
}
\begin{document}

\maketitle
 
\begin{abstract}

Aphasia is a language disorder that affects the speaking ability of millions of patients. This paper presents a new benchmark for Aphasia speech recognition and detection tasks using state-of-the-art speech recognition techniques with the AphsiaBank dataset. Specifically, we introduce two multi-task learning methods based on the CTC/Attention architecture to perform both tasks simultaneously. Our system achieves state-of-the-art speaker-level detection accuracy (97.3\%), and a relative WER reduction of 11\% for moderate Aphasia patients. In addition, we demonstrate the generalizability of our approach by applying it to another disordered speech database, the DementiaBank Pitt corpus. We will make our all-in-one recipes and pre-trained model publicly available to facilitate reproducibility. Our standardized data preprocessing pipeline and open-source recipes enable researchers to compare results directly, promoting progress in disordered speech processing.


\end{abstract}

\section{Introduction}\label{sec:intro}

Aphasia is a language disorder that affects patients' abilities to communicate effectively.
This condition can manifest in various components of the language, including phonology, grammar, and semantics, among others~\cite{broca_aphasia,speech_errors_aphasia}.
Recent studies have developed machine learning methods for Aphasia speech recognition and detection to assist clinicians in the diagnosis and documentation process.
The recognition task involves transcribing Aphasia speech into text, while the detection task requires classifying whether a speaker has Aphasia.

For the recognition task, various automatic speech recognition (ASR) architectures have been benchmarked on Aphasia speech data.
A recent trend is using a pre-trained Wav2vec2.0~\cite{wav2vec2} to perform zero-shot or few-shot predictions for low-resource languages~\cite{aphasia_zeroshot,aphasia_spanish}.
Other benchmarked ASR models include DNN-HMM~\cite{DucLe2016,aphasia_asr_moe} and RNN~\cite{DucLe2018,QinYing2020,QinYing2019}.
While some studies formulate the detection task as a binary classification problem~\cite{aphasia_zeroshot,Aparna2019}, others consider it as an Aphasia Quotient prediction task~\cite{DucLe2018,QinYing2020,QinYing2018,Qin2018BLSTM}.
Aphasia Quotient (AQ) is a metric used to measure the severity of Aphasia~\cite{WAB}.
Linguistic statistics extracted from transcripts or ASR output are commonly used as input features.
They include filler words per minute, pauses to words ratio, number of phones per word, and many more~\cite{aphasia_zeroshot,DucLe2018,Aparna2019,QinYing2018,Qin2018BLSTM,QinYing2020}.
Some researchers incorporate acoustic information as well since it also contains signs of Aphasia~\cite{QinYing2020,QinYing2019,QinYing2018}.
The classification or regression models used in these studies vary from classical machine learning models such as SVM~\cite{aphasia_zeroshot,DucLe2018,Aparna2019,QinYing2018} to deep neural networks~\cite{QinYing2019,Qin2018BLSTM,Qin2022,Dunfield2020}.

Although several ASR systems have been tested in these studies, we believe performance can be further improved by leveraging recent state-of-the-art ASR architectures. Furthermore, as most existing Aphasia detectors require text as the input, an ASR system is required if the transcription is not available.
Since ASR errors can cascade into the detection system, the detection accuracy might be suboptimal.
Therefore, we aim to build an end-to-end system that can perform both tasks simultaneously using the latest ASR technologies.
This system should be able to derive linguistic features from acoustic input implicitly and utilize both of them for the tasks.

To the best of our knowledge, we are the first to present an architecture that can detect the presence of Aphasia on both the sentence and the speaker level, while simultaneously transcribing the speech to text.
Our system has two variants and achieves state-of-the-art detection performance on the AphasiaBank English subset.
This is achieved with the help of the hybrid CTC/Attention ASR architecture~\cite{hybrid_ctc_attention}, E-Branchformer~\cite{EBF}, and WavLM~\cite{WavLM}.
Among existing studies, we found inconsistencies in evaluation metrics, data compositions, and preprocessing procedures.
Therefore, we make our code and pretrained model open-source in the hope of establishing a standardized benchmark environment for both tasks\footnote{\url{https://github.com/espnet/espnet}}.
We demonstrate the effectiveness and generalizability of our approach by applying it to another disordered speech database, the DementiaBank Pitt corpus~\cite{DementiaBankPitt}.


\section{Method}\label{sec: method}

In this section, we present a system that jointly models Aphasia detection and Aphasia speech recognition.
The techniques used in this system have all been proven to be state-of-the-art in various speech processing tasks~\cite{hybrid_ctc_attention,EBF,superb,multilingual_interctc_hier}.

\subsection{Hybrid CTC/Attention}\label{sec:hybrid_ctc_att}

%

Our proposed method is based on the hybrid CTC/Attention ASR architecture~\cite{hybrid_ctc_attention}.
This architecture comprises an encoder, denoted by $\Enc(\cdot)$, and a decoder, denoted by $\Dec(\cdot)$.
The encoder captures the acoustic information and can optionally generate a text sequence using Connectionist Temporal Classification (CTC)~\cite{CTC}.
The text sequence is primarily predicted by an attention-based decoder in an auto-regressive manner given the encoder's hidden states~\cite{hybrid_ctc_attention}.

The input to the encoder, denoted as $\mathbf{X} = (\mathbf{x}_l \in \mathbb{R}^D | l=1,\dots,L)$, is a sequence of $L$ acoustic feature vectors, where each vector has $D$ dimensions.
The ground truth text sequence is denoted as $T = (t_k \in V | k=1,\dots,K)$, which contains $K$ text tokens from a vocabulary $V$.
Using the CTC algorithm~\cite{CTC}, the encoder predicts the likelihood of generating the text sequence given the input $P_{\Enc}(T|X)$.
The encoder hidden state output is denoted as $H$:
\begin{align}
     \mathbf{H} &= \Enc(\mathbf{X}) \label{eq:encoder_output} \\
     P(T|X)     &= \CTC(\mathbf{H})
\end{align}
The decoder models $P(T|X)$ given the encoder hidden states and prior token predictions~\cite{povey16_interspeech}:
\begin{align}
P(t_k|X,T_{1:k-1}) &= \Dec(\mathbf{H}, T_{1:k-1}) \\
P_{\Dec}(T|X) &\approx \prod_{k}^K P(t_k|X,T_{1:k-1})
\end{align}
During training, the model is optimized using the weighted sum of the CTC loss and the decoder loss~\cite{hybrid_ctc_attention}:
\begin{align}
    \mathcal{L} &= \lambda\mathcal{L}_{\CTC} + (1-\lambda)\mathcal{L}_{\Dec} \label{eq:asr_loss} \\
    &= -\lambda\log P_{\Enc}(T|X) - (1-\lambda)\log P_{\Dec}(T|X)
\end{align}
where the CTC weight $\lambda$ is an hyper-parameter.
The output of the encoder and decoder is jointly decoded using beam search to produce the final hypothesis during inference~\cite{hybrid_ctc_attention}.
The system is often evaluated with word error rate (WER).

\subsection{Intermediate CTC}\label{sec:interctc}

Intermediate CTC (InterCTC) was proposed to regularize deep encoder networks and to support multi-task learning~\cite{interctc,interctc_self_condition,multilingual_interctc_hier}.
To achieve this, the existing CTC module is applied to the output of an intermediate encoder layer with index $e$.
Then subsequent encoder layers incorporate the intermediate predictions into their input.
Equation~\ref{eq:encoder_output} can be reorganized as:
\begin{align}
    \mathbf{H}_e &= \Enc_{1:e}(\mathbf{X}) \\
    P(Z_{\text{Inter}}|X)     &= \CTC(\mathbf{H}_e) \\
    \mathbf{H} &= \Enc_{e+1:E}( \Norm(\mathbf{H}_e) + \Linear(Z_{\text{Inter}}) )
\end{align}
where $E$ is the total number of encoder layers, and $Z_{\text{Inter}}$ is the latent sequence of the InterCTC target sequence $T_{\text{Inter}} = (t'_k | k=1,\dots,K')$.
$\Norm(\cdot)$ and $\Linear(\cdot)$ refer to a normalization layer and a linear layer respectively.
The negative log likelihood of generating $T_{\text{Inter}}$ is used as the InterCTC loss:
\begin{align}
    \mathcal{L}_{\text{Inter}} &= -\log P_{\text{Inter}}(T_{\text{Inter}}|X)
\end{align}
The choice of $T_{\text{Inter}}$ is dependent on the task.
During training, the intermediate layer is optimized to correctly predict $T_{\text{Inter}}$ by including $\mathcal{L}_{\text{Inter}}$ in the loss function:
\begin{align}
    \mathcal{L}_{\CTC}' &= \alpha \mathcal{L}_{\text{Inter}}
    + (1-\alpha) \mathcal{L}_{\CTC} \label{eq:new_ctc_loss}
\end{align}
where the InterCTC weight $\alpha$ is a hyper-parameter.
The updated overall loss function is obtained by inserting Equation~\ref{eq:new_ctc_loss} into Equation~\ref{eq:asr_loss}:
\begin{align}
    \mathcal{L}' &= \lambda\mathcal{L}_{\CTC}' + (1-\lambda)\mathcal{L}_{\Dec}
\end{align}
Note that it is possible to apply CTC to multiple encoder layers while having different target sequences for each.
In that case, the average of all InterCTC losses is used as $\mathcal{L}_{\text{Inter}}$~\cite{interctc_self_condition,multilingual_interctc_hier}.

\subsection{Speech Recognizer}\label{sec:method_aphasia_asr}

Our speech recognizer follows the design of a hybrid CTC/Attention architecture described in Section~\ref{sec:hybrid_ctc_att}.
It transcribes the acoustic feature sequence $\mathbf{X}_{ij}$ belonging to speaker $s_j$ to the corresponding text token sequence $T_{ij}$.

We experiment with recently proposed encoder architectures that enhance acoustic modeling ability over the original Transformer. 
One of these architectures, called Conformer, sequentially combines convolution and self-attention.
This allows for capturing both the global and the local context of input sequences~\cite{conformer}.
On the other hand, Branchformer models these contexts using parallel branches and merges them together.
Both architectures demonstrate competitive performance in speech processing tasks~\cite{Branchformer}.
In subsequent studies, E-Branchformer is proposed to enhance Branchformer further.
It comprises a better method for merging the branches, and it achieves the new state-of-the-art ASR performance~\cite{EBF}.

Meanwhile, self-supervised learning representation (SSLR) has been developed to improve the generalizability of acoustic feature extraction.
SSLR leverages a large amount of unlabeled speech data to learn a universal representation from speech signals.
Studies show significant performance improvement in ASR and other downstream tasks by using  SSLR as the input of the encoder~\cite{HuBERT,WavLM,superb,E2EIntegrationSSL}.

\subsection{Aphasia Detectors}\label{sec:method_detectors}

We present two types of Aphasia detectors based on the speech recognizer, the tag-based detector and the InterCTC-based detector.
Inspired by the use of language identifiers in multilingual speech recognition~\cite{multilingual_asr_e2e,multilingual_lid,multilingual_lid_transformer},
we form an extended vocabulary $V'$ by adding two Aphasia tag tokens to $V$:
\begin{align}
    V' = V \cup \{ \texttt{[APH]}, \texttt{[NONAPH]} \}
\end{align}
We then train the ASR model using $T_{ij}^{\text{tag}} = (t_k\in V'|k=1,\dots,K)$ as the ground truth, where $T_{ij}^{\text{tag}}$ is formed by inserting one or more Aphasia tags to $T_{ij}$.
Specifically, \texttt{[APH]} is inserted if the speaker has Aphasia while \texttt{[NONAPH]} is inserted if the speaker is healthy.
This method effectively trains the model to perform both tasks jointly.
Moreover, the model leverages both linguistic and acoustic information to detect Aphasia, as the encoder first generates an initial tag prediction based on the acoustic features, and the decoder then refines the prediction based on prior textual context.
During inference, the sentence-level prediction is obtained by taking out the tag token from the predicted sequence.
Three tag insertion strategies will be tested in Section~\ref{sec:experiments}: prepending, appending, and using both.
We note that all tag tokens are excluded from WER computation.

InterCTC is proven to be effective at identifying language identity in multilingual ASR as part of a multi-task objective.
By conditioning on its language identity predictions, the ASR model achieves state-of-the-art performance on FLEURS~\cite{multilingual_interctc_hier}.
Inspired by this, the second type of Aphasia detector uses InterCTC to classify input speech as either Aphasia or healthy speech.
During training, the ground truth sequence $T_{ij}^{\text{inter}}$ for InterCTC contains an Aphasia tag token.
During inference, the prediction $\hat{y}_{ij}$ is generated by checking the tag produced by InterCTC greedy search.
This approach allows us to select which encoder layer to use for the best speaker-level accuracy.

For both the tag-based and InterCTC-based detectors, the speaker-level Aphasia prediction $y_j$ is obtained via majority voting of $y_{ij}$ for all $i$.

\section{Experiments}\label{sec:experiments}

In this section, we first explore the impact of state-of-the-art encoder architectures and SSLR on Aphasia speech recognition.
We then analyze the performance of the proposed method for recognition and detection tasks.
All of our experiments were conducted using ESPnet2~\cite{espnet}.

\subsection{Datasets}

\subsubsection{CHAT Transcripts}\label{sec:CHAT}

CHAT~\cite{CHILDES} is a standardized transcription format for describing conversational interactions, used by both AphasiaBank and DementiaBank.
Besides the textual representations of spoken words, it includes a set of notations that describes non-speech sounds, paraphasias, phonology, morphology, and more.
The transcript cleaning procedures differ between prior works, making it difficult to fairly compare their machine learning systems.
Therefore, we derive a pipeline based on previous work~\cite{aphasia_spanish} in the hope of standardizing this process for future research.

The specific steps of our pipeline are as follows.
(1)~Keep the textual representations of retracing, repetitions, filler words, phonological fragments, and IPA annotations while removing their markers.
(2)~Replace laughter markers with a special token~\texttt{<LAU>}.
(3)~Remove pre-codes, postcodes, punctuations, comments, explanations, special utterance terminators, and special form markers
(4)~Remove markers of word errors, interruption, paralinguistics, pauses, overlap precedes, local events, gestures, and unrecognized words.
(5)~Remove all empty sentences after the above steps.

\subsubsection{AphasiaBank and DementiaBank}

AphasiaBank~\cite{aphasiabank} is a popular speech corpus among the existing work.
The dataset contains spontaneous conversations between investigators and Aphasia patients.
It also includes conversations with healthy individuals as the control group.
All experiments in this paper are performed using the English subset.
Similar to~\cite{aphasia_spanish}, we obtain the training, validation, and test set by drawing 56\%, 19\%, and 25\% percent of Aphasic speakers from each severity.
There are four severity levels, each corresponding to a range of AQ scores: mild (AQ $>$ 75), moderate (50 $<$ AQ $\le$ 75), severe (25 $<$ AQ $\le$ 50), and very severe (0 $\le$ AQ $\le$ 25)~\cite{aphasiabank}.
The control group is split using the same ratio and merged with patients' data.
Doing so ensures our data splits are representative across all severity levels.
We then slice the recordings into sentences using the timestamps provided in the CHAT transcripts while cleaning them as described in Section~\ref{sec:CHAT}. 
After that, sentences shorter than $0.3$ seconds or longer than $30$ seconds are removed.
Before data augmentation, the training set contains $42.7$ hours of patient data and $22.7$ hours of control group data while the test data contains $20.1$ and $10.1$ hours.
Details can be found in our code release.

Dementia speech recognition and detection have been a popular research topic as well~\cite{ConformerDementia,Hu2022ExploitingCA,Ammar2019,Bertini2022,Bhat22019,Chen2019}.
We use the DementiaBank Pitt corpus~\cite{DementiaBankPitt} to test the generalizability of our design.
Similar to recent studies~\cite{ConformerDementia,Hu2022ExploitingCA}, we use the ADReSS challenge~\cite{ADReSS2020} test set, which is a subset of the DementiaBank Pitt corpus, for evaluation and the remaining data in the corpus for training and validation.
We note that audio from the challenge test set has been enhanced with noise removal and volume normalization, while the transcripts have been preprocessed.
To preserve a consistent data pipeline, we instead use the original recordings and transcripts from the Pitt corpus as our test data.
Details can be found in our code base.

\subsection{Experimental Setups}

\noindent\textbf{Baseline:}
We first build two ASR systems using Conformer~\cite{conformer} and E-Branchformer~\cite{EBF}, as described in Section~\ref{sec:method_aphasia_asr}.
The Conformer encoder has $12$ blocks, each having $2048$ hidden units and $4$ attention heads.
The E-Branchformer encoder has $12$ blocks, each with $1024$ hidden units, and $4$ attention heads.
The cgMLP module has $3072$ units and the convolution kernel size is $31$.
Both systems use a Transformer decoder
with $6$ blocks, each having $2048$ hidden units and $4$ attention heads.
The Conformer and E-Branchformer models have $44.2$ and $45.7$ million trainable parameters respectively. 
For the detection task, we reproduce the Aphasia detection experiment from a previous study.
The detector is a support vector machine (SVM) that takes in linguistic features extracted from the oracle transcript to predict a binary classification label~\cite{aphasia_zeroshot}.

\noindent\textbf{Proposed Method:}
We first build a system with learned acoustic representations extracted from WavLM~\cite{WavLM} as the input to the E-Branchformer encoder.
Using it as a foundation, we build tag-based and InterCTC-based detectors as described in Section~\ref{sec:method_detectors}.
We also investigate the impact of tag insertion positions: prepending, appending, and both.
Meanwhile, we apply InterCTC to the 6th and the 9th encoder layer respectively, and analyze their performance difference.
We set both the CTC and InterCTC weight to $0.3$ and the inference beam size to $10$.

In all experiments, we use speed perturbation with ratios of $0.9$ and $1.1$, as well as SpecAugment~\cite{SpecAug}, to augment the data.
We choose the Adam optimizer with a learning rate of $10^{-3}$ and a weight decay of $10^{-6}$.
We employ \texttt{warmuplr} learning rate scheduler with $2500$ warm-up steps and a gradient clipping of $1$.
Each final model is selected by averaging the $10$ checkpoints with the highest validation accuracy out of $40$ epochs.
More details can be found in our code base.

\begin{table}[t]

\centering \resizebox {0.9\linewidth} {!} {
\begin{threeparttable}
\begin{tabular}{lccc}
\toprule 
\multirow{2}{*}{Model} & Patient & Control  & Overall \\
& WER & WER & WER \\
\midrule

\textit{Baselines} &  &  & \\
Conformer  & $40.3$ & $35.3$ & $38.1$ \\
E-Branchformer & $36.2$ & $31.2$ & $34.0$ \\
\midrule

\textit{Proposed Methods} &  &  & \\
E-Branchformer+WavLM      & $26.4$ & $17.0$ & $22.2$ \\
+Tag-prepend              & $26.3$ & $16.9$ & $22.2$ \\
+Tag-append               & $\mathbf{26.2}$ & $16.9$ & $\mathbf{22.1}$ \\
+Tag-prepend/append       & $26.3$ & $\mathbf{16.8}$ & $\mathbf{22.1}$ \\
+InterCTC-6               & $26.3$ & $16.9$ & $\mathbf{22.1}$ \\
+InterCTC-9               & $26.3$ & $16.9$ & $22.2$ \\
+InterCTC-6/Tag-prepend   & $26.3$ & $16.9$ & $\mathbf{22.1}$ \\
\bottomrule
\end{tabular}

\end{threeparttable}
}

\caption{\label{tab:aphasia_asr_results} Word error rate (WER) of proposed methods evaluated on AphasiaBank. }
\vspace{-8mm}
\end{table}

\subsection{Results and Discussion}

Overall, the proposed systems achieve both accurate Aphasia speech recognition and detection at the same time.
As shown in Table~\ref{tab:aphasia_asr_results}, switching from Conformer to E-Branchformer leads to a significant ASR performance improvement by $4.1$ WER absolute.
Adding WavLM reduces the WER further by $11.8$.
This proves the effectiveness of using a state-of-the-art ASR encoder and SSLR for Aphasia speech recognition.
Surprisingly, both types of detectors lead to a slightly better ASR performance than the vanilla ASR model ($0.1$ WER reduction).
This implies that the ASR predictions can be refined based on Aphasia detection results.
We compare the ASR performance of our systems with previous work in detail in Table~\ref{tab:aphasia_asr_comparisons}.
Our systems obtained significant lower WER for mild, moderate, and severe patients, even against systems using an external language model.
Despite this, they have a much higher WER for very severe Aphasia patients.
We believe this is because hybrid CTC/Attention architectures are data-hungry, but the number of utterances and their average duration is much smaller for very severe patients.

\begin{table*}

\centering \resizebox {0.85\linewidth} {!} {
\begin{threeparttable}
\begin{tabular}{lcccccccc}
\toprule 
Model & Metric & \multicolumn{5}{c}{Patient}  & Control  & Overall \\ 
      &        & Overall & Mild & Moderate & Severe & Very severe &          &         \\
\midrule
DNN-HMM~\cite{DucLe2016} & PER & - & $47.4$ & $52.8$ & $61.0$ & $75.8$ & - & -  \\
DNN-HMM + MOE~\cite{Perez2020} & PER & $36.8$ & $33.1$ & $41.6$ & \multicolumn{2}{c}{\cellcolor{gray!20}$62.9$} & - & - \\
Wav2vec2 (zero-shot)~\cite{aphasia_zeroshot} & WER & $56.0$ & - & - & - & - & $37.5$ & $47.1$ \\
BLSTM-RNN+i-Vector+LM~\cite{DucLe2018} & WER & - & $33.7$ & $41.1$ & $49.2$ & $63.2$ & - & - \\
Wav2vec2~\cite{aphasia_spanish} & WER & - & $23.6$ & $36.8$ & $36.4$ & $\mathbf{59.1}$ & - & - \\

\midrule
\multicolumn{9}{l}{\textit{E-Branchformer+WavLM}} \\
+Tag-prepend    & WER & $\mathbf{26.3}$ & $22.3$ & $32.8$ & $\mathbf{34.5}$ & $72.5$ & $\mathbf{16.9}$ & $22.2$ \\
+InterCTC-6     & WER & $\mathbf{26.3}$ & $22.3$ & $\mathbf{32.6}$ & $34.7$ & $71.7$ & $\mathbf{16.9}$ & $\mathbf{22.1}$ \\
+InterCTC-6/Tag-prepend & WER & $\mathbf{26.3}$ & $\mathbf{22.1}$ & $32.9$ & $34.8$ & $73.3$ & $\mathbf{16.9}$ & $\mathbf{22.1}$ \\
\bottomrule
\end{tabular}
\end{threeparttable}
}

\caption{\label{tab:aphasia_asr_comparisons} The recognition word error rate of proposed methods and existing work on the AphasiaBank English subset. The metrics are phoneme error rate (PER) and word error rate (WER). Note that existing studies use different data splits than ours. }
\vspace{-6mm}
\end{table*}

\begin{table}[t]

\centering 
\begin{threeparttable}
\begin{tabular}{lcc}
\toprule 
\multirow{2}{*}{Model} & \multicolumn{2}{c}{Accuracy} \\ 
& Sent & Spk \\
\midrule
SVM~\cite{aphasia_zeroshot}  & -      & $96.2$ \\
\midrule
\multicolumn{3}{l}{\textit{E-Branchformer+WavLM}} \\
+Tag-prepend                 & $89.3$ & $95.1$ \\
+Tag-append                  & $89.2$ & $95.1$ \\
+Tag-prepend/append          & $\mathbf{90.8}$ & $95.7$ \\
+InterCTC-6                  & $85.2$ & $\mathbf{97.3}$ \\
+InterCTC-9                  & $84.5$ & $\mathbf{97.3}$ \\
+InterCTC-6/Tag-prepend      & $89.7$ & $96.7$ \\
\bottomrule
\end{tabular}

\end{threeparttable}

\caption{\label{tab:aphasia_detection_results} Sentence-level (Sent) and speaker-level (Spk) detection accuracy of proposed methods on AphasiaBank.
\cite{aphasia_zeroshot} is reproduced using the official code with oracle transcripts as the input.
For +Tag-prepend/append and +InterCTC-6/Tag-prepend experiments, only the Tag-prepend output is reported since the difference is negligible.
}
\vspace{-8mm}
\end{table}

From Table~\ref{tab:aphasia_detection_results}, we can see that the tag-based Aphasia detectors have the best sentence-level Aphasia detection accuracy.
Interestingly, although the performance difference between prepending and appending Aphasia tags is insignificant, inserting at both positions leads to slightly better sentence-level and speaker-level accuracy.
Meanwhile, the InterCTC-based detector at layer $6$ achieves state-of-the-art speaker-level accuracy ($97.3\%$), surpassing the SVM baseline.
However, its sentence-level accuracy is lower than those of tag-based detectors.
This corresponds to previous studies showing that middle encoder layers are more important to speaker-related tasks while the bottom layers are more relevant to ASR and related tasks~\cite{WavLM,E2EIntegrationSSL}.
We also find that tag-based detectors produce significantly more false positives for speakers who do not have Aphasia but are less fluent than others, thus having a lower speaker-level accuracy.
This implies that tag-based detectors are sometimes too sensitive to dysfluency.

Finally, more accurate tag-based predictions can be obtained by combining InterCTC and tag-prepending.
This suggests that tag predictions are refined based on prior InterCTC predictions.
A similar result is discovered in a previous study where the language identity predictions are more accurate by incorporating an InterCTC auxiliary task~\cite{multilingual_interctc_hier}.
In addition, the combined model has higher sentence-level accuracy and lower speaker-level accuracy compared to its InterCTC counterpart, which demands future investigation.

\begin{table}

\centering \resizebox {0.98\linewidth} {!} {
\begin{threeparttable}
\begin{tabular}{lccccc}
\toprule 
\multirow{2}{*}{Model} & \multirow{2}{*}{Patient} & \multirow{2}{*}{Control} & \multirow{2}{*}{Overall} & \multicolumn{2}{c}{Accuracy} \\ 
& & & & Sent & Spk \\
\midrule
Conformer~\cite{Hu2022ExploitingCA} & - & - & $29.7$  & - & - \\
Conformer~\cite{ConformerDementia} & - & - & $25.5$  & - & $91.7$  \\
\midrule
\multicolumn{3}{l}{\textit{E-Branchformer+WavLM}} \\
+Tag-prepend & $39.1$ & $15.0$ & $\mathbf{24.8}$  & $65.6$ & $83.3$ \\
+InterCTC-6  & $39.6$ & $15.0$ & $25.1$  & $61.3$ & $77.1$ \\
\bottomrule
\end{tabular}

\end{threeparttable}
}

\caption{\label{tab:dementia_results} Test result of proposed methods on DementiaBank.
The metric for speech recognition is the word error rate (WER). The metrics for Dementia detection are sentence-level (Sent) and speaker-level (Spk) accuracy.
Other studies~\cite{Ammar2019,Bertini2022,Bhat22019,Chen2019} are not listed as their models are trained and tested on different data.
Note that \cite{Hu2022ExploitingCA,ConformerDementia} use a larger and cleaner training set.}

\vspace{-8mm}
\end{table}

Table~\ref{tab:dementia_results} shows evaluation results for DementiaBank.
Although the overall WER is much lower than those in previous studies, Dementia detection accuracy is suboptimal.
As we drew original recordings from the DementiaBank Pitt corpus, the audio is often noisy and has variable speaking volume.
Consequently, the model is less effective at acoustic modeling, as seen by the decreased InterCTC detection accuracy.
The results also suggest that linguistic features are more important for Dementia detection than Aphasia.
Furthermore, majority voting for speaker-level predictions is less effective in this case as the number of sentences per speaker is typically between 5 to 20.
Despite this, we believe our method has the potential to be adapted to other disordered speech in future studies.

\section{Conclusion}\label{sec: conclusion}


In this paper, we build an all-in-one Aphasia speech recognition and detection system and test its performance using AphasiaBank and DementiaBank.
We also standardize the data processing and model evaluation process to establish a public benchmark.
Future studies are required to improve the recognition performance for severe Aphasia patients and the detection performance on DementiaBank.
We can also further investigate the impact of joint learning and combining detector methods, and explore the potential benefits of fine-tuning a pre-trained healthy ASR system using disordered speech.

\section{Acknowledgements}
This work used the Bridges2 system at PSC and Delta system at NCSA through allocation CIS210014 from the Advanced Cyberinfrastructure Coordination Ecosystem: Services \& Support (ACCESS) program, which is supported by National Science Foundation grants \#2138259, \#2138286, \#2138307, \#2137603, and \#2138296.


\section{References}
{
\printbibliography
}

\end{document}